\documentstyle[12pt,psfig]{article}
\newcommand{\NP}[1]{ Nucl.\ Phys.\ {\bf #1}}

\newcommand{\PL}[1]{ Phys.\ Lett.\ {\bf #1}}

\newcommand{\PR}[1]{Phys.\ Rev.\ {\bf #1}}
\newcommand{\PRL}[1]{ Phys.\ Rev.\ Lett.\ {\bf #1}}

\begin{document}
\begin{flushright}
\end{flushright}
\vskip3.cm
\begin{center}
{\LARGE {Theoretical Approaches
to HERA Physics}\footnote{Contribution to the Proceedings
of the Madagascar High-Energy
Physics International Conference (HEP-MAD'01), Antananarivo (Madagascar),
27 September-5 October, 2001, edited by Stephan Narison, World Scientific
(to be published).}}\\[2ex]
C. Merino\\
{\small\em Department of Particle Physics, University of Santiago           
de Compostela}\\
{\small\em 15706 Santiago de Compostela, Galiza, Spain}\\ 
{\small\em e-mail: Merino@fpaxp1.usc.es}
\end{center}

\begin{center}
Abstract
\end{center}
A review is presented of the different theoretical 
models proposed to approach consistently 
the interplay between soft and hard physics, 
that can now be studied experimentally at HERA for the first time.

\section{HERA Physics}

The range of the physics scanned by HERA is very broad, from hadroproduction 
to jets, charm,
quark fragmentation and instantons, even though in this contribution we 
will only 
focus on low $x$.

The setting up of HERA, the first electron-proton collider, at DESY during
the nineties of the last century made possible the experimental study of deep
inelastic scattering (DIS) processes and, more generally, of quantum 
chromodynamics (QCD), under kinematical conditions where the interplay between
soft and hard physics should play an important role~\cite{kuhlen}. 

The most important measurement in DIS is that of the cross-section of the 
process:

\centerline{$ep\rightarrow eX$,}

\noindent  
as a function of any pair of independent Lorentz invariants built from the 
kinematic variables $q$, the four-momentum transfer mediated by the virtual 
photon, and $P$,
the four-momentum of the incoming proton,
e.g. 
$(Q^2,x)$:
\begin{equation}
Q^2=-q^2,\hskip0.2cm x=\frac{Q^2}{2Pq},
\label{eq:variables}              
\end{equation}

The large center of mass energy $s$ provided by the $ep$ collider allows 
the detectors of the two international experimental collaborations working 
at HERA, H1~\cite{h1} and ZEUS~\cite{zeus}, the study of the kinematic 
regimes both at large
$Q^2$ and very small $x$. This newly accessible region of very small $x$
is specially interesting from the theoretical point of view. Since 
\begin{equation}
x=\frac{Q^2}{Q^2+s+m^2_p},
\label{eq:s}
\end{equation}
at fixed $Q^2$ the limit $x\ll 1$ corresponds to the limit of large $s$, 
$s\gg Q^2$ (Regge limit), where the Regge Field Theory, which was used to
describe the hadronic processes before QCD was accepted as the general
field theory accounting for the strong interaction, should be valid. 
Thus, HERA makes possible
the study of DIS where both the perturbative QCD limit, $Q^2\gg\Lambda^2$,
and the Regge limit should apply, i.e. where the hard-perturbative and
soft-nonperturbative physics should interplay and shed light on the fundamental
question of confinement. On the other hand, very small $x$ means
parton densities in which the proton momentum fraction is being shared by many
partons (mainly gluons), i.e. very high density parton densities, and thus
this region of very small $x$ connects the study of DIS with the physics
of the heavy-ion collisions, so important in the next future with the 
operative start of the Large Hadron Collider (LHC) at CERN.    

Theoretically, the DIS differential cross-section can be expressed in terms
of two independent structure functions $F_1(x,Q^2)$ and $F_2(x,Q^2)$:
\begin{equation}
\frac{d^2\sigma}{dxdQ^2}=\frac{4\pi\alpha^2}{xQ^4}
\left[\left(1-y+\frac{y^2}{2}\right) F_2
-\frac{y^2}{2} F_L\right],
\label{eq:dsigma}
\end{equation}
where $\alpha$ is the electromagnetic coupling constant, $y$ is the fraction
of the electron energy tranferred to the proton in the proton rest frame
$(0<y<1)$, and $F_L(x,Q^2)$ is the longitudinal structure function:
\begin{equation}
F_L=F_2-2xF_1.
\label{eq:fl}
\end{equation}
The extraction of $F_2$ from the cross-section measurement~(\ref{eq:dsigma})
implies an assumption on $F_L$, since there is no direct $F_L$ measurements
in the HERA regime. Once this assumption is made, we can compare the 
experimental
$F_2$ with the theoretical one, that can be written 
in terms of the quark and antiquark densities, $q_i$ and $\bar q_i$,
and the quark charges, $e_{q_i}$. However, it has not been possible
to derive the hadronic structure from first principles including the 
interactions of quarks and gluons as given by QCD, and the same
happens when calculating the $ep$ cross-section. 

In fact, only three things are rigorously calculable in QCD~\cite{yndurain},
due to the fact that they involve just one large scale:
the process $e^+e^-\rightarrow hadrons$, $F_2$ in DIS and the form factor
in exclusive processes~\cite{brodsky}. In DIS, 
the equations that include the perturbative effects and give the relation 
between the parton distribution function
taken at two different scales are called evolution equations, and they
can be obtained rigorously by extracting the perturbative part of $\sigma_{ep}$
in the Operator Product Expansion (OPE) frame and summing up the large 
logarithmic perturbative corrections in the Renormalization Group (RG) 
equations~\cite{gw} for the regime with a large scale:
\begin{equation}
Q^2\rightarrow\infty,\hskip0.2cm with\hskip0.2cm 
\frac{Q^2}{2\nu}=x\sim 1\hskip0.2cm (\alpha_s lnx\ll 1).
\label{eq:apregime}
\end{equation}
An equivalent formulation to the one based on OPE
but expressed in terms of parton language is the one of the DGLAP 
equations~\cite{dglap}. The physical meaning of this method is very clear
since it deals with parton densities and fragmentation functions, which
are basic quantities for describing short-distance reactions~\cite{muta}.
From (\ref{eq:apregime}), we see that the DGLAP equations can be written 
by neglecting the terms in $ln\left(Q^2/s\right)$ at
the leading log approximation in $ln\left(Q^2/Q^2_0\right)$.
This leads to consider all ladder diagrams in which the transverse momenta 
are strongly ordered: 
\begin{equation}
Q^2\ll\cdots k^2_{T_i}\ll k^2_{T_{i+1}}\ll\cdots Q^2.
\label{eq:qorder}
\end{equation}    
This is expected to be good enough when $Q^2$ is large but $x$ is not too
small. In this approximation the DGLAP evolution equation for the parton density 
$q_i$ corresponding to the quark of
flavor $i$ is (a similar equation can be written for the 
evolution of the gluon density, $g$):
\begin{equation}
\frac{dq_i(x,Q^2)}{dlnQ^2}=\frac{\alpha_s}{2\pi}
\int_x^1\frac{dz}{z}\left[q_i(z,Q^2)P_{qq}\left(\frac{x}{z}\right)
+g(z,Q^2)P_{qg}\left(\frac{x}{z}\right)\right].
\label{eq:dglap}
\end{equation}
The $P\left(\frac{x}{z}\right)$ functions in~(\ref{eq:dglap}) are the 
calculable splitting functions giving the probabilities for the parton 
branchings.

When $x$ is small but $Q^2$ is not very large the 
DGLAP equations can no longer be used. It is for this new regime:
\begin{equation}
\Lambda^2\ll Q^2\ll s,
\label{eq:bfklregime}
\end{equation}
where no theoretically rigorous 
evolution equation is available, that a new evolution equation analytically
solvable at fixed
$\alpha_s$ has been proposed, the BFKL evolution equation~\cite{bfkl}, 
in which we can, from (\ref{eq:bfklregime}), 
consider the leading log
approximation in $ln\left(1/x\right)$, which translates in resumming
the ladder diagrams 
with strongly ordered $x_i$:
\begin{equation}
x_0\ll\cdots x_i\ll x_{i+1}\ll\cdots x,
\label{eq:xorder1}
\end{equation}   
and no ordering on $k_{T_i}$. The BFKL equation is an evolution equation in $x$
of the gluon density $g(x,k_T^2)$, dominant at small $x$:
\begin{equation}
\frac{\partial g(x,k_T^2)}{\partial ln(1/x)}=\frac{3\alpha_s}{\pi}
\int_0^{\infty}\frac{dk^{\prime 2}_T}{k^{\prime 2}_T}
\left[\frac{g(x,k^{\prime 2}_T)-g(x,k^2_T)}{\vert k^{\prime 2}_T-k^2_T\vert}
+\frac{g(x,k^2_T)}{\sqrt{4k^{\prime 2}_T-k^2_T}}\right].
\label{eq:bfkl}
\end{equation} 
Thus $g(x,k^2_T)$ can be calculated for any small $x$ once it is known at some
starting value $x_0$, for all $k^2_T$.

As already mentioned, the BFKL equation can be solved analytically 
for fixed $\alpha_s$, with the
$x$ and $k_T$ behaviors of the solution being:
\begin{equation}
g(x,k^2_T)\propto \left(\frac{x}{x_0}\right)^{\lambda}
\sqrt{k^2_T}\hskip0.1cm\frac{1}{ln 1/x},
\label{eq:lpom}
\end{equation} 
with 
\begin{equation}
\lambda^{LO}=(n_c\alpha_s/\pi) 4\hskip0.1cm ln2 \sim 0.5 
\label{eq:lpom1}
\end{equation}
for $n_c=3$ colors,
being the famous BFKL Pomeron intercept. At this point one has to note
that different problems appear when trying to solve the BFKL equation
at the next-to-leading order (NLO)~\cite{nlobfkl}, among them the impossibility 
of treating 
all the terms in the new solution as Regge terms. Still, the NLO estimates of
the BFKL Pomeron intercept seem to lead to a value:
\begin{equation}
\lambda^{NLO}\sim 0.25\div 0.3.
\label{eq:nlolpom}
\end{equation}
Also attempts have been made to include the running of $\alpha_s(Q^2)$ in
the calculation.

In the region where $x$ is small and $Q^2$ is not too large, now accessible 
at HERA, the old Regge Field Theory should work.
The Regge Field Theory~\cite{rft} is a method 
based on analyticity, crossing symmetries between the $s$ and $t$ channels 
and unitarity in the complex angular momenta plane, valid to compute the
cross-section of hadron-hadron collisions at high energies (the
equivalent in DIS to small $x$), by squaring the sum over the scattering
amplitudes due to the mesons that can be exchanged in the process. These
mesons can be empirically grouped along the so called Regge trajectories, 
parametrized as straight lines which relate the mass $m$ and the spin $J$
of the exchanged mesons, $J=\alpha(m^2)$:
\begin{equation}
\alpha(t)=\alpha_0+\alpha^{\prime}t,
\label{eq:rtraject}
\end{equation}
where $\alpha_0$ is the intercept and $\alpha^{\prime}$ the slope of the 
trajectory. Thus the elastic cross-section for a hadron-hadron collision
will be calculated by summing over all the Regge trajectories whose resonances
can be exchanged in the reaction, and the final result can be written as:
\begin{equation}
\frac{d\sigma^{el}}{dt}\propto (\beta(t))^2 s^{2\alpha(t)-2},
\label{eq:sigmael}
\end{equation}
with $\beta(t)$ an unknown real function.
Now, by using the optical theorem which relates the total cross-section to the 
forward elastic amplitude, we can predict the behavior of the total
hadron-hadron cross-section:
\begin{equation}
\sigma_{tot}\propto s^{\alpha_0-1}.
\label{eq:sigmatot}
\end{equation}
The huge amount of experimental data on hadronic cross-sections for many
different processes shows an universal and steady rise at large energies
that can only be accounted for by parametrizing all these cross-sections
as the sum of two different components:
\begin{equation}
\sigma_{tot}=As^{\alpha_{R}(0)-1}+Bs^{\alpha_{P}(0)-1},
\label{eq:sigpar}
\end{equation}
with $A$ and $B$ process-dependent constants and the intercepts 
$\alpha_{R}(0)\sim 0.5$ and $\alpha_{P}(0)\sim 1.08$, universal 
process-independent constants. While the first term in
Eq.~(\ref{eq:sigpar}) represents the exchange of the experimentally
detected mesonic resonances (secondary Reggeon), the second term
represents the exchange of a hypothetical object, the soft Pomeron, which has
the vacuum quantum numbers (electrically and color neutral, isospin $0$ and
$C$-parity $+1$), and it is the one responsible, through the value of its 
intercept larger than $1$ (supercritical Pomeron),
for the rise of $\sigma^{tot}$
at large energies. 
This soft Pomeron, as it also happens for 
the BFKL Pomeron, is suspected to be of gluonic nature.
Of course, Equation~(\ref{eq:sigpar}) violates the unitarity constraints
imposed by the Froissart bound on the cross-section behavior, so we must 
interpret the soft Pomeron just as an useful phenomenological tool.

As a matter of fact, DIS processes at small $x$ can be viewed in terms of the 
Regge Field Theory as virtual photon-proton scattering at high energy, with
$F_2$ dominated by the gluon content of the proton:
\begin{equation}
F_2=-\frac{Q^2}{4\pi^2\alpha}\sigma^{tot}_{\gamma^* p}
\rightarrow F_2\propto\left(1/x\right)^{\alpha_P(0)},
\label{eq:DISregge}
\end{equation}
where this should be compared with the solution of the BFKL evolution
equation, so we can write:
\begin{equation}
F_2\propto\left(1/x\right)^{\alpha_{BFKL}},
\label{eq:DISbfkl}
\end{equation}                        
and we view the exchange of a hard BFKL Pomeron in the Regge language
as the sum over graphs with one gluon ladder between the interaction particles
in perturbative QCD.

Now we are in conditions to address the problems at the origin of the
theoretical models that we pretend to review in this contribution. The
first question is how to connect QCD and the Regge Field Theory in a
theory which could compute quantitatively any DIS or hadronic process
in both the soft and hard regimes. 
The second problem concerns the perturbative QCD prediction
of a strong increase of the parton densities at low $x$, which led to the
idea~\cite{glr} that the density of partons (gluons) becomes so big that at 
some point these gluons
cannot be considered as independent partons any more, but they interact
among them. To determine whether this is actually the case, and whether these
interactions must be taken into account for a consistent description of DIS
processes in this small $x$ regime, is what is called the saturation problem, 
at present
intensively analyzed, both theoretically and experimentally at HERA.
To precisely establish~\cite{mueller1} the way saturation relates to the 
idea of unitarity, 
a better understanding
of the fundamental question of confinement is needed.

\section{Theoretical Approaches}

Many models have been proposed in the attempt to make the connection between 
soft and hard physics in a theoretically consistent way during the last years.
Given the current lack of both theoretical and experimental tools needed in
order to articulate an universal and fundamental approach, all these models 
are phenomenological 
to a larger or smaller degree, and they can basically
be classified in three main goups. The first class of models explicitly
includes saturation, as the dipole models which use the idea
that the virtual photon actually splits into a $q$-$\bar{q}$ dipole and
it is this dipole which subsequently interacts with the proton. Here, the size 
of the dipole will determine the soft or hard character of the interaction. 
The second class of models, that we will call phenomenological 
parametrizations of 
structure functions at low $Q^2$, provides a parametrization of the structure
function $F_2(x,Q^2)$ describing the experimental data in the 
(nonperturbative) 
region where perturbative QCD is supposed to fail, and then uses 
this parametrization
to be plugged in as an initial condition
in the perturbative evolution equations to
obtain a description of the experiment in the whole kinematic range. The
philosophy behind the third group of models (dynamical models for the low $Q^2$
behavior of $F_2$) is just the opposite. Now a QCD-based consistent 
description of the data in the perturbative region is considered, and then
extrapolated down to the region of not large $Q^2$,
through the evolution equations and under certain
ad-hoc conditions imposed to maintain the consistency of the description 
in the region
where the perturbative treatment is supposed not to work any more.

The first explicit realization of the idea of saturation was presented not for
the nucleon, but for
the nuclear case~\cite{mueller2}, when it was shown that the Glauber model 
without gluon interaction leads to the saturation of the $1/k^2_T$ growth 
of the quark and gluon distributions at fixed $s$ as $k^2_T$ becomes small.

One dipole model very transparent in its physical
interpretation is the GW model~\cite{gbw}. Here the structure function $F_2$ is 
separated into two terms: 
\begin{equation}
F_2=F^T+F^L,
\label{gw1}
\end{equation}
with, for $x\ll 1$:
\begin{equation}
F^{T,L}=\frac{Q^2}{4\pi^2\alpha}\int d^2\vec{r}dz
\vert\Psi^{T,L}(\vec{r},z,Q^2)\vert^2
\hat{\sigma}(x,\vec{r}).
\label{gw2}
\end{equation}
Here $\Psi^{T,L}$ is the known wave function for a transverse (T) or 
longitudinal (L) polarized virtual photon $\gamma^*$ to split into a 
$q$-$\bar{q}$ dipole,
$\hat{\sigma}$ is the dipole cross-section describing the interaction
of the dipole with the proton, and $\vec{r}$ is the transverse separation of 
the $q$-$\bar{q}$ pair. Unitarity is built in by the phenomenological form 
of the dipole cross-section:
\begin{equation}
\begin{array}{rcl}
\hat{\sigma}&=&\sigma_0\left(1-exp\left(-r^2/4R^2_0(x)\right)\right) \\[4pt]
R_0(x)&=&\left(1/Q_0\right) (x/x_0)^{\Delta/2},
\label{gw3}
\end{array}
\end{equation}
where $R_0$ is the so-called saturation radius, $Q_0=1.\hskip0.1cm GeV$, 
and parameters
$\sigma_0$, $x_0$ and $\Delta$ are fitted to all inclusive DIS data with 
$x<0.01$. Thus in this model, at small $r$ one has color transparency and
a strong growth of $\hat{\sigma}$ with $x$:
\begin{equation}
\hat{\sigma}\sim r^2 x^{-\Delta},
\label{gw4}
\end{equation}
while at large $r$ (or $x\rightarrow 0$) $\hat{\sigma}$ approaches the 
black-disk constant value $\sigma_0$ (saturation). The transition to saturation
is governed by $R_0(x)$.

There are models including other saturation mechanisms. One of them is
the interaction between partons in the parton cascade~\cite{levin},
not taken into account in QCD evolution equations, but that could become 
important
to slow down the growth of parton densities. Here, the parton interactions will 
create an equilibrium-like system of partons with a definite value for the average 
transverse momentum, $Q_s(x)$ (saturation scale). A different proposed
saturation mechanism is the
percolation of strings~\cite{as}, a second order phase transition which takes 
place when clusters of overlapping strings, with size of the order of the total
transverse area available, appear. This phase transition is used as an indication 
for the onset of saturation
of the density of partons. In this approach, which can be generalized to the nuclear 
case, a multiple exchange model for $ep$ collisions is needed.

For the nuclear case, the first approach~\cite{mueller2} presenting the
$k^2_T$ saturation curves for the quark and gluon distributions at fixed $s$ has
been extended~\cite{ab} to the case where interaction among gluons is taken into 
account, obtaining the $x$ dependence of those curves. Also all multiple Pomeron
LO exchanges have been included~\cite{kovch} in deriving a small $x$ evolution 
equation of $F_2$ for a large nucleus from the first nuclear approach without 
interaction~\cite{mueller2}, and it is shown that in the double leading log limit 
this equation reduces to the GLR equation~\cite{glr}.
In a different  
model~\cite{mcv}, the valence quarks of the nucleons of the nucleus
are treated as the sources of the small $x$ gluon distribution of the nucleus.

Among the phenomenological parametrizations of structure functions at 
low $Q^2$, the DL model~\cite{dl} is a parametrization of $F_2$ that uses
two separate (soft and hard) Pomerons:
\begin{equation}
\begin{array}{rcl}
F_2(x,Q^2)&=&f_0(q^2) x^{-\epsilon_0}(1-x)^7\\[4pt]
&+&f_1(q^2) x^{-\epsilon_1}(1-x)^7\\[4pt]
&+&f_2(q^2) x^{-\epsilon_2}(1-x)^3,
\label{dl1}
\end{array}
\end{equation}
with $\epsilon_0=0.4372$ (hard), $\epsilon_1=0.0808$ (soft), and
$\epsilon_2=-0.4525$ (valence). The ABY model~\cite{aby} is another
parametrization of $F_2$ using two different components, a hard Pomeron
plus a soft Pomeron that in this model is taken as a flat term.

The CKMT model~\cite{ckmt} is also a phenomenological parametrization
of $F_2$ at low $Q^2$, but it uses only one effective Pomeron. The CKMT model
proposes for the nucleon structure functions:
\begin{equation}
F_2(x,Q^2) = F_S(x,Q^2) + F_{NS}(x,Q^2),
\label{eq:eq1}
\end{equation}
the following parametrization of its two terms in
the region of small and moderate $Q^2$. For the singlet term, corresponding to
the Pomeron contribution:
\begin{equation}
F_S(x,Q^2) = A^{S}\hskip0.1cm x^{-\Delta(Q^2)}(1-x)^{n(Q^2)+4}
\left({Q^2\over Q^2+a}\right)^{1+\Delta(Q^2)},
\label{eq:eq2}
\end{equation}
where the $x$$\rightarrow$0
behavior is determined by an effective intercept
of the Pomeron,~$\Delta$,
which takes into account Pomeron cuts and, therefore (and this is one of the
main points of the model), it depends on $Q^2$. This dependence is
parametrized
as :
\begin{equation}
\Delta (Q^2) = \Delta_0\left(1+{\Delta_1 Q^2
\over Q^2+\Delta_2}\right).
\label{eq:eq3}
\end{equation}
Thus, for low values of $Q^2$ (large cuts), $\Delta$ is close
to the effective value found from analysis of hadronic total cross-sections
($\Delta$$\sim$0.08), while for high values of $Q^2$ (small cuts),
$\Delta$ takes the bare Pomeron value,
$\Delta$$\sim0.2\div0.25$. The
parametrization for the non-singlet term, which corresponds to the secondary
Reggeon (f, $A_2$) contribution, is:
\begin{equation}
F_{NS}(x,Q^2) = B^{NS}\hskip0.1cm x^{1-\alpha_R}(1-x)^{n(Q^2)}
\left({Q^2\over Q^2+b}\right)^{\alpha_R},
\label{eq:eq4}
\end{equation}
where the $x$$\rightarrow$0 behavior is determined by the secondary
Reggeon intercept $\alpha_R$, which is in the range $\alpha_R\sim0.4\div0.5$.
The valence quark contribution can be separated into the contribution of the
u ($B^{NS}_u$) and d ($B^{NS}_d$) valence quarks,
the normalization condition for valence quarks fixes
both contributions
at one
given value of $Q^2$ ($Q_v^2=2.\hskip0.1cm GeV^2$ has been used in the calculations).
For both the singlet and the non-singlet terms, the behavior when
$x$$\rightarrow$1 is                                                            
controlled by $n(Q^2)$, with $n(Q^2)$ being
\begin{equation}
n(Q^2) = {3\over2}\left(1+{ Q^2
\over Q^2+c}\right).
\label{eq:eq5}
\end{equation}
Therefore, for $Q^2=0.$ the behavior of the valence quark distributions
is given by
Regge intercepts, $n$(0)=$\alpha_R$(0)$-$$\alpha_N$(0)$\sim$ 3/2, while
the behavior of
$n(Q^2)$ for large $Q^2$ is taken to coincide with
dimensional counting rules.
The total cross-section for real ($Q^2$=0) photons can be obtained from the
structure function $F_2$ using the following relation:
\begin{equation}
\sigma^{tot}_{\gamma p}(\nu) = \left[{4\pi^2\alpha\over Q^2}
F_2(x,Q^2)\right]_{Q^2=0}.
\label{eq:eq6}
\end{equation}
The proper $F_2(x,Q^2)$$\sim$$Q^2$
behavior when
$Q^2$$\rightarrow$0, is given in the
model
by the last factors in Equations (\ref{eq:eq2}) and (\ref{eq:eq4}), leading
to the following form of the $\sigma^{tot}_{\gamma p}(\nu)$ in
the CKMT model:
\begin{equation}
\sigma^{tot}_{\gamma p}(\nu) = 4\pi^2\alpha
\left(A^{S}\hskip0.1cm a^{-1-\Delta_0}(2m\nu)^{\Delta_0}
+(B^{NS}_u+B^{NS}_d)\hskip0.1cm b^{-\alpha_R}(2m\nu)^{\alpha_R-1}\right).
\label{eq:eq6a}
\end{equation}
The parameters in the model were determined from a joint
fit of the
$\sigma^{tot}_{\gamma p}$ data and NMC data on
the proton structure function in the region
$1.\hskip0.1cm GeV^2 \leq Q^2 \leq 5.\hskip0.1cm GeV^2$, obtaining a very good
description of the available experimental data.                                   
The next step in this approach is to introduce the QCD evolution in
the partonic distributions of the CKMT model and thus to
determine the structure
functions at higher values of $Q^2$. For this, the
evolution equation in two loops in the $\overline{\mbox{MS}}$
scheme with
$\Lambda=200.\hskip0.1cm MeV$ was used.
The results obtained by taking into account the QCD evolution in
this way are in a very good agreement with
the experimental data on
$F_2(x,Q^2)$ at high values of $Q^2$.                                           

The ALLM parametrization~\cite{allm} of $F_2$ also uses $Q^2$-dependent powers 
of $x$, but it does no introduce QCD evolution.

Among the dynamical models of the low $Q^2$ behavior of $F_2$, we should mention
the GRV model~\cite{grv}, in which the QCD evolution equations are extended down 
to the very low $Q^2$ region ($Q^2<1.\hskip0.1cm GeV^2$) using dynamical 
parton densities 
generated radiatively from valence-like inputs at some resolution scale. In this
model both LO and NLO approximations are used. Another dynamical model is the
KP model~\cite{kp}, where the DGLAP evolution equation at NLO is solved by giving
analytical parametrizations for the parton ditributions. This model takes into
account the contributions of higher-twist (renormalon-type) operators of the Wilson
OPE, which are important at low $Q^2$. Finally, the BK model~\cite{bk} considers
the contributions both from the parton model with QCD corrections extended to the 
low $Q^2$ region and from the low mass vector mesons.     
There are other vector meson dominance (VMD) 
models~\cite{mrs}.  

On top of these, one can find models which are in the middle of two of the
groups above. Thus, in the NZZ model~\cite{nzz} a color dipole approach is used
to solve the BFKL equation with  running coupling constant, while another 
dipole model with both a hard and a soft Pomeron has been proposed~\cite{dd} where
the large dipoles couple to the soft Pomeron and small dipoles couple to the
hard Pomeron. This model has been applied to the case of the charm structure
function $F_2^c(x,Q^2)$. Also an explicit dipole model with the CKMT pattern
of energy behavior (effective $Q^2$-dependent Pomeron intercept) has recently
been presented~\cite{cfks}. Other attempts have been made by interpolating
between Regge behavior and the high $Q^2$ DGLAP asymptotics~\cite{jenk}. 

\section{Discussion and Conclusions}

In this contribution we have presented a basic introduction to HERA physics, mainly
focused on the field of low $x$ physics.
The low $x$ HERA physics provides an important experimental tool to obtain a
better understanding of the interplay between soft and hard physics and the relation
between concepts as saturation and unitarity, and to
address more consistently the fundamental question of confinement.                    
Models including saturation present a decrease of the effective Pomeron
intercept as $x\rightarrow 0$, while in other models the effective intercept
should increase as energy (or $1/x$) increases.

The experimental evidence of saturation at HERA, specifically the
presence of a change in the $Q^2$ dependence of $F_2$ at very small $x$ and
moderate $Q^2$, presented in a very graphic way as the turnovers of the 
logarithmic slopes of $F_2$, in particular of 
$\partial F_2/\partial lnQ^2$ at moderate $Q^2$ and small $x$, is not
conclusive. 

Essential information on the behavior of the structure functions in the region
of extremely small $x$, not accessible at HERA, will be available at LHC
in a hopefully near future.
                                      
\section*{Acknowledgments}
It is a pleasure to thank and congratulate Stephan Narison for organizing this 
conference and for his effort to connect, in a two-way profitable relation,
our physics community to such a great country as
Madagascar and to its scientists. 
I am also grateful to Marie Razafindrakoto for her success in implementing
a friendly organization for the participants before and during the conference,
and in dealing with the proceedings.
Thanks are due to Fy Rafam'andrianjafy for her help on the week-end tour
to the National Park of Andasibe and Manambato, and to all members of 
the local committee. 
I wish to thank M.A. Braun and G. Parente for reading and improving 
the manuscript, and E.G. Ferreiro for her useful comments.
Finally, I would like to express my hope that this conference will become 
only the first of a long and fruitful series to come.

\end{document}